\documentclass[sigconf]{acmart}
\usepackage{wrapfig}
\usepackage{pifont}
\usepackage{multirow}
\usepackage{adjustbox}
\usepackage{listings}
\usepackage[font=footnotesize]{caption}
\usepackage{subcaption}
\usepackage{enumerate}
\usepackage{xspace}

\definecolor{NL-description}{RGB}{128, 0, 128}
\definecolor{model-output}{RGB}{0, 80, 0}
\definecolor{instance-specific-input}{RGB}{0, 0, 128}
\definecolor{instance-inspecific-input}{RGB}{0, 0, 0}

\newcommand{\numberOfProjects}{9}

\newcommand{\numberOfMUT}{18}
\newcommand{\numberOfTestsPerMUT}{two}
\newcommand{\numberOfQueriesPerTemperature}{10}
\newcommand{\numberOfCandidateTests}{100}

\newcommand{\para}[1]{\smallskip\noindent {\bf #1} }

\newcommand{\code}[1]{{\ttfamily \small #1}}

\AtBeginDocument{%
  \providecommand\BibTeX{{%
    \normalfont B\kern-0.5em{\scshape i\kern-0.25em b}\kern-0.8em\TeX}}}

\setcopyright{none}
\renewcommand\footnotetextcopyrightpermission[1]{} 

\hypersetup{draft}

\begin{document}

\title[A Study of Few-Shot, Pre-Trained Language Models on Code]{Code Generation Tools (Almost) for Free?\\ A Study of Few-Shot, Pre-Trained Language Models on Code}

\author{Patrick Barei\ss}
\affiliation{%
  \institution{University of Stuttgart}
  \country{Germany}}
\author{Beatriz Souza}
\affiliation{%
	\institution{Federal University of Pernambuco}
	\country{Brasil}}

\author{Marcelo d'Amorim}
\affiliation{%
	\institution{Federal University of Pernambuco}
	\country{Brasil}}

\author{Michael Pradel}
\affiliation{%
	\institution{University of Stuttgart}
	\country{Germany}}


\begin{abstract}
Few-shot learning with large-scale, pre-trained language models is a powerful way to answer questions about code, e.g., how to complete a given code example, or even generate code snippets from scratch.
The success of these models raises the question whether they could serve as a basis for building a wide range code generation tools.
Traditionally, such tools are built manually and separately for each task.
Instead, few-shot learning may allow to obtain different tools from a single pre-trained language model by simply providing a few examples or a natural language description of the expected tool behavior.
This paper studies to what extent a state-of-the-art, pre-trained language model of code, Codex, may serve this purpose.
We consider three code manipulation and code generation tasks targeted by a range of traditional tools:
(i) code mutation;
(ii) test oracle generation from natural language documentation; and
(iii) test case generation.
For each task, we compare few-shot learning to a manually built tool.
Our results show that the model-based tools complement (code mutation), are on par (test oracle generation), or even outperform their respective traditionally built tool (test case generation), while imposing far less effort to develop them.
By comparing the effectiveness of different variants of the model-based tools, we provide insights on how to design an appropriate input (``prompt'') to the model and what influence the size of the model has. For example, we find that providing a small natural language description of the code generation task is an easy way to improve predictions.
Overall, we conclude that few-shot language models are surprisingly effective, yet there is still more work to be done, such as exploring more diverse ways of prompting and tackling even more involved tasks.
\end{abstract}


\maketitle

\section{Introduction}

Various software engineering tools assist developers by generating source code.
One group of approaches reasons about existing code and modifies it in a way suitable to achieve some goal.
For example, code mutation tools~\cite{jia2010analysis,papadakis2019mutation} introduce mistakes to measure the effectiveness of test suites, and automated program repair tools~\cite{cacm2019-program-repair,monperrus2018automatic} suggest how to fix programming mistakes.
Another group of approaches generates new code from scratch, given some existing code that the new code is supposed to relate to.
For example, test case generators~\cite{Csallner2004,Pacheco2007,Fraser2011a} automatically create tests that exercise a given method under test, and code completion tools~\cite{Bruch2009,Raychev2014,Hellendoorn2019} generate code that completes an existing code snippet in a suitable way.
Finally, a third group of code generation tools does not require any existing code as an input, but instead generates new code given some natural language artifact.
For example, some approaches generate test oracles based on informal API documentation~\cite{Goffi2016,Blasi2018,Blasi2021}, infer API usage protocols~\cite{Zhong2009a}, or suggest missing type annotations~\cite{icse2019}.

The traditional way of creating such code manipulation tools is based on program analysis combined with various rules and heuristics.
Program analysis can, at least in principle, ensure that the generated code is guaranteed to have certain properties, e.g., to be type-correct or to pass a given set of test cases.
Hand-coded rules and heuristics are typically required to enable a technique to be effective and efficient on real-world software.
More recently, learning-based approaches have started to complement traditional program analysis-based code generation tools~\cite{NeuralSoftwareAnalysis}.
Typically, these approaches formulate the specific code generation task as a supervised learning problem, and require large amounts of training data to obtain an effective machine learning model.
A commonality of both traditional program analyses and learning-based approaches is that creating a new code generation tool involves significant human effort.
Even worse, this effort often must be repeated for each new combination of a task to achieve and a programming language to target.

A recent trend in the natural language processing (NLP) community promises a form of ``general intelligence'' that remedies many of the problems of building task-specific techniques:
\emph{few-shot learning with large-scale, pre-trained language models}~\cite{Brown2020}, henceforth abbreviated with \emph{FSLMs}.
These models are trained on huge amounts of data without focusing on a specific downstream task.
Instead, the training is based on generic pseudo-tasks for which it is trivial to obtain sufficient training data, e.g., predicting masked words or whether two sentences belong together.
Once trained, FSLMs are effective at various question answering and text generation tasks, e.g., reading comprehension, trivia quizzes, translation between languages, and text completion~\cite{Brown2020}.

Applying FSLMs to code is still a relatively sparsely explored area.
While recent work employs pre-training of models of code as a means to reduce the amount of required training examples~\cite{Ahmad2021,Guo2021,Liu2020,Feng2020}, these approaches still fine-tune a model for a specific purpose and hence require moderately large amounts of labeled training examples.
Noteworthy exceptions include GitHub's Copilot code completion system\footnote{\url{https://copilot.github.com/}}, which is based on the Codex FSLM~\cite{Chen2021}, and the recently released, open-source PolyCoder model family~\cite{Xu2022}.
While the results of these models are impressive, code completion is only one of many code generation tasks.
Do the abilities of FSLMs generalize to other software engineering tasks that traditionally have been addressed by special-purpose code generation techniques?
In case of a positive answer, FSLMs offer the potential to obtain code generation tools (almost) for free, as an FSLM gets trained once and can then be applied to many different tasks.
Despite this potential and the strong interest of the software engineering community in automated code generation techniques, there currently is no systematic study of the abilities of FSLMs on such tasks.

\begin{table*}[t!]
    \centering
    \caption{Representative examples of results obtained with FSLM-based code generation tools.}
    \vspace{-1ex}
    \begin{adjustbox}{width=\textwidth}
      \begin{tabular}{p{3cm}p{5cm}p{5cm}}
\toprule
Task & Example \#1 & Example \#2 \\
\midrule
\multirow{3.5}{*}{Code mutation}  & 
\begin{lstlisting}[language=Java,linewidth=8.5cm,basicstyle=\ttfamily\footnotesize,breaklines=true,escapechar={@},belowskip=-1em,aboveskip=-0.5em]
parsed = (parsed + "000000000").substring(0, 9);
|==>
parsed = (parsed + "000000").substring(0, 9);
\end{lstlisting}
& 
\begin{lstlisting}[language=Java,linewidth=8.5cm,basicstyle=\ttfamily\footnotesize,breaklines=true,escapechar={@},belowskip=-1em,aboveskip=-0.5em]
ofSeconds(divRem[0].longValue(), divRem[1].intValue())
|==@\textgreater@
ofSeconds(divRem[0].longValue() / divRem[1].intValue(), divRem[1].intValue())
\end{lstlisting}
\\
\midrule
\multirow{7}{*}{Oracle generation} & 
\begin{lstlisting}[language=Java,linewidth=8.5cm,basicstyle=\ttfamily\footnotesize,keywordstyle==\ttfamily\footnotesize,breaklines=true,escapechar={@},belowskip=-1em,aboveskip=-0.5em]
The returned implementation is immutable, thread-safe and
Serializable. It is equivalent to tick(system(zone), Duration.ofSeconds(1)).
|==@\textgreater@
java.time.Clock.tick(java.time.Clock.system(zone),
  java.time.Duration.ofSeconds(1))
IS EQUIVALENT TO 
java.time.Clock.tickSeconds(zone)
\end{lstlisting}
& 
\begin{lstlisting}[language=Java,linewidth=8.5cm,basicstyle=\ttfamily\footnotesize,keywordstyle==\ttfamily\footnotesize,breaklines=true,escapechar={@},belowskip=-1em,aboveskip=-0.5em]
Note for Java 7 and later: this method should be treated
as deprecated; use the equivalent Long#compare method instead.
|==@\textgreater@
com.google.common.primitives.Longs.compare(a, b)
IS EQUIVALENT TO
java.lang.Long.compare(a, b)
\end{lstlisting}
\\
\midrule
\multirow{9}{*}{Test generation}   & 
\begin{lstlisting}[language=Java,linewidth=8.5cm,basicstyle=\ttfamily\footnotesize,breaklines=true,escapechar={@},belowskip=-1em,aboveskip=-0.5em]
public Vector1D scalarMultiply(double a) {
  return new Vector1D(a * x); }
|==@\textgreater@
public void testScalarMultiply() {
  Vector1D v1 = new Vector1D(1);
  Vector1D v2 = new Vector1D(2);
  Assert.assertEquals(v1.scalarMultiply(2), v2); }
\end{lstlisting}
& 
\begin{lstlisting}[language=Java,linewidth=8.5cm,basicstyle=\ttfamily\footnotesize,breaklines=true,escapechar={@},belowskip=-1em,aboveskip=-0.5em]
public boolean contains(Object o) {
  return indexOf(o) >= 0; }
|==@\textgreater@
public void testContains() {
  ArrayList<String> list = new ArrayList<String>();
  list.add("a");
  list.add("b");
  list.add("c");
  Assert.assertTrue(list.contains("b"));
  Assert.assertFalse(list.contains("d")); }
\end{lstlisting}\\
\bottomrule
\end{tabular}

      \label{fig:cool_examples}
    \end{adjustbox}
\end{table*}

This paper presents the first systematic study of FSLMs as the key ingredient for creating code generation tools.
We describe a general framework for creating a code generation tool based on an existing FSLM, apply it to three popular tasks that are representative for different kinds of code generation problems, and compare the FSLM-based approach against traditionally developed state-of-the-art tools.
Instantiating our framework for a specific code generation tasks involves three steps.
First, develop an extractor of code or natural information to use in a query to the model.
Second, design a suitable \emph{prompt}, i.e., a template of how to present the input to the model, which then gets instantiated for each given example.
Finally, develop a lightweight post-processing module, which, e.g., removes generated code that fails to compile.
We argue that these steps are lightweight compared to designing and implementing a traditional program generation technique, as they leave the most challenging parts of the tasks to the FSLM.
As a result, the approach offers an almost-for-free way of obtaining a code generation tool.

We instantiate these ideas for three code generation tasks: \emph{code mutation}, \emph{test oracle generation}, and \emph{test case generation}.
These tasks have received significant interest from the software engineering community, and hence, offer state-of-the-art tools to compare against.
The tasks also cover different levels of granularity of the generated code, ranging from manipulating a few tokens in code mutation to generating entire test cases.
Finally, the selected tasks are based on different kinds of input: code mutation and test case generation are based on existing code, whereas test oracle generation is based on natural language documentation. Table~\ref{fig:cool_examples} shows two representative example outputs that FSLM-based tools produce for each of these tasks. The examples follow the format $x|==>y$, where $x$ and $y$ denote, respectively, the input and output of the prompt for the given task.

For each task, we instantiate our general framework to create an FSLM-based code generation tool and then apply the tool to real-world software.
We then systematically compare the results produced by the FSLM-based tool against an existing, state-of-the-art tool built specifically for the same purpose:
the Major~\cite{major} code mutation tool,
the MeMo~\cite{Blasi2021} test oracle extraction tool,
and the Randoop~\cite{Pacheco2007} test case generator.
We measure the effectiveness of each tool using metrics of success suitable for the task, e.g., code coverage for test case generation, and precision/recall w.r.t.\ a ground truth for test oracle generation.

Our key findings include:
\begin{itemize}
\item FSLM-based tools are similarly and sometimes even more effective than existing, special-purpose tools.
For example, for oracle generation, we measure an F1 score of 0.59 and 0.60 for MeMo~\cite{Blasi2021} and an FSLM-based tool, respectively.
For test generation, Randoop achieves 10\% coverage, whereas a simple FSLM-based tool achieves 14\%.
\item FSLM-based and traditionally-developed tools often complement each other.
For example, our FSLM-based code mutation tool creates various mutants that Major cannot generate.
The complementary nature of the two kinds of tools shows the potential of combining traditional and FSLM-based approaches.
For example, combining Randoop-generated and FSLM-generated test cases yields 16\% coverage, i.e., it exceeds both approaches individually.
\item FSLM-based tool do not come completely for free.
To be effective, they need specifically designed prompts and suitable inputs extracted from the given code or natural language.
Yet, the effort required to create an FSLM-based tool is clearly lower than that for building special-purpose code generation tools from scratch.
\end{itemize}

In summary, this paper contributes the following:
\begin{itemize}
\item The first systematic study of FSLM-based code generation tools.
\item We are the first to address code mutation, test oracle generation, and test case generation in an end-to-end manner with general-purpose FSLMs.
\item Insights that show the potential and challenges of building FSLM-based code generation tools, providing guidance for future work.
\end{itemize}

\section{Background}


A \emph{generative language model} is designed to predict the next token given some previous tokens. For example, if such a model is given the input ``I am Barack Obama. I used to be the president of the United States of'', such a language model might predict ``America'' as the next token. This can be used to generate text by repeatedly sampling for the next token.
When using such a model for downstream tasks that differ from the next token prediction objective, the step of initial training is often referred to as \emph{pre-training}.

A pre-trained model can be adapted to a specific downstream task via \emph{fine-tuning}, i.e., in additional training step based on labeled data for the downstream task.
A recently proposed alternative is \emph{few-shot learning}~\cite{Brown2020}, which refers to the ability to perform a task without any fine-tuning, but given only very few (typically, between one and ten) examples as part of the query to the model.
We utilize generative language models as few-shot learners, which we refer to as \emph{few-shot learning with large-scale, pre-trained language models} (FSLM). 
We use OpenAI's Codex~\cite{Chen2021} model, which is trained on a large set of GitHub projects.
We access the model through its API. Alternative generative models exist, e.g., GPT-NeoX~\cite{black2022gpt}.



The input provided to an FSLM is referred to as the \emph{prompt}.
Prompts typically contain a few examples of inputs with their desired outputs, followed by the input for which the model should provide an answer.
For the above example, a prompt could start by giving a few example pairs of head of states and the corresponding country, and then ``Barack Obama'', to which the model might respond with ``United States''.
Prompts are, in principle, unstructured text, and what exactly is provided in a prompt may strongly influence the results.
When querying an FSLM with a prompt, the user can select the \emph{temperature}, which intuitively speaking controls the creativity or randomness of the model's responses.
A higher temperature means the model will generate more diverse responses, but be less factual and precise.
Repeatedly querying a model may return different results, especially when using a higher temperature.

\section{Methodology}


\begin{figure*}
    \centering
    \includegraphics[width=.8\linewidth]{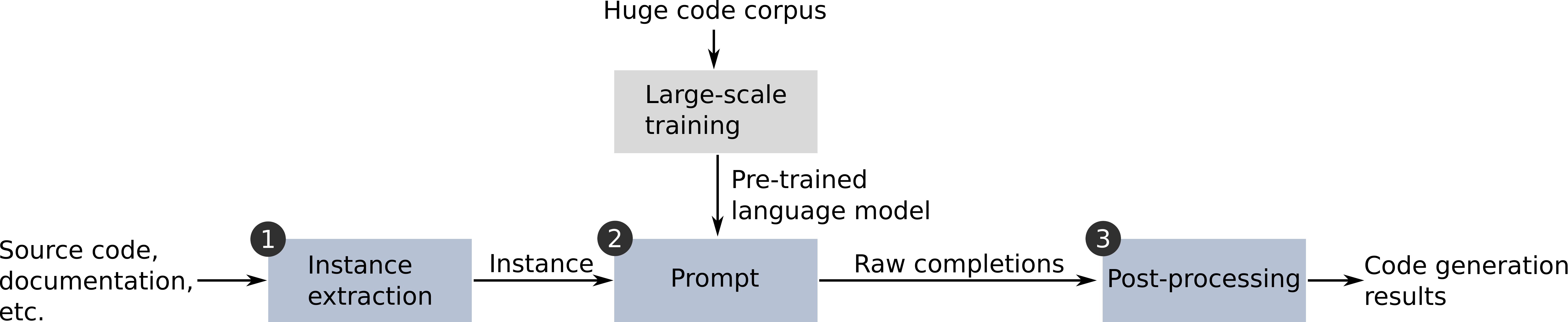}
    \caption{Overview of a general framework for generating code analysis tools using few-shot, pre-trained language models.}
    \label{fig:approach_overview}
\end{figure*}



Figure~\ref{fig:approach_overview} shows a general framework for producing code generation tools for a diverse set of tasks.
The framework relies on a large-scale language model pre-trained on code, such as Codex~\cite{Chen2021}.
The input to the framework is a textual representation of a
software artifact, e.g., source code or documentation.
The output is a set of generated code snippets, e.g., a modified version of the given source code, an executable specification, or a test case.
The framework is organized in three main steps, which we briefly describe in the
following.

\begin{enumerate}
\item \textbf{Instance extraction.}
The first step is responsible for extracting parts of a given software artifact that are relevant for the code generation task.
We refer to an extracted part as an \emph{instance}.
For example, for code mutation, the instance extraction takes in source code and extracts code lines for which we want to generate mutants. The rationale for not simply passing in the entire raw software artifact is two-fold.
First, FSLMs impose a maximum input size, e.g., 4,096 tokens for the Codex model series.
Second, larger inputs take longer to process, i.e., the instance extraction reduces the overall time to generate code.

\item \textbf{Prompt design.}
The second step to use our framework is designing an effective prompt, which is perhaps the most difficult part of creating an FSLM-based code generation tool.
The prompt contains (i) an instance, as extracted in the previous step, and (ii) contextual information, such as examples for addressing the code generation task and/or a natural language description of the task.
The prompts we use in our study include a part that is invariant across all instances (e.g., a natural language description of the task) and a part that is instance-specific (e.g., the line of code to mutate).
Given a prompt for a specific instance, the approach passes the prompt to the FSLM and then obtains a completion of it.

\item \textbf{Post-processing.}
Finally, the third step is to post-process the raw output produced by the model in order to obtain the final code generation results.
The post-processing may filter the completions, e.g., to ensure that the generated code compiles or to copy the predicted code into a task-specific code template.

\end{enumerate}

Sections~\ref{sec:task-one},~\ref{sec:task-two}, and~\ref{sec:task-three} describe the code generation tasks that this paper focuses on according to the three above steps.

\subsection{Research Questions}
\label{sec:rqs}

The overall goal of this study is to understand the strengths and limitations of FSML-based code generation tools.
To this end, we investigate the following research questions.

\begin{enumerate}[RQ1.]

\item \para{Accuracy:} How accurate are the model's predictions compared to existing tools?

\item \para{Impact of Prompt:} What kinds of prompts are most effective at producing accurate results?

\item \para{Impact of Model Size:} How much does the size of the FSLM influence the accuracy?

\end{enumerate}

The motivation for RQ1 is that building traditional tools by hand imposes significant human costs.
Understanding to what extent a single general-purpose language model could replace these tools may help reducing the cost for creating new tools.
The motivation for RQ2 is that the prompt to query a pre-trained model is the main ``knob'' to control the quality of the model's predictions.
Understanding what prompts are effective (or not) helps in making best use of the existing models.
Finally, the motivation for RQ3 is that state-of-the-art language models are trained on huge datasets using enormous computational resources.
Understanding the impact of model size on the model's effectiveness will help appropriately allocate computational resources to train models.

\subsection{\label{sec:task-one}Task 1: Code Mutation}

We address our research questions by studying them on three popular code generation tasks.
The first task is code mutation, a popular technique to assess the quality of a test suite by estimating its ability to detect injected faults. Code mutation modifies a given piece of code by injecting a programming mistake.
As a simple example, a code mutation tool may change a comparison \code{x > 5} into \code{x < 5}.


\subsubsection{Baseline Tool}

We study the effectiveness of an FSLM-based code mutation tool by comparing it against Major~\cite{major}, a popular code mutation tool for Java.
Major applies different built-in mutation operators and ensures that all created mutants compile.

\subsubsection{Instance Extraction}
To create an FSLM-based code mutation tool, the first step is extracting code snippets to modify.
Since mutation operators typically are local code transformations, an instance for this task consists of a single line of code.
The instance extractor takes a Java file as input and returns a list of lines of code that we then try to mutate via the FSLM.
For a fair comparison, we focus our experiments on those lines where Major applies a mutation. 

\begin{figure}[tb]
    \begin{lstlisting}[language=Java,linewidth=8.5cm,basicstyle=\ttfamily\footnotesize,breaklines=true,otherkeywords={describe, it},escapechar={@}]
@\color{NL-description}Generate mutations for the following snippets of code.@
@\color{instance-inspecific-input}[[Code]]@
@\color{instance-inspecific-input}long biasedExp = (longBits \& DoubleConsts.EXP\_BIT\_MASK)
    >> (DoubleConsts.SIGNIFICAND\_WIDTH - 1);@
@\color{instance-inspecific-input}[[Mutations]]@
@\color{instance-inspecific-input}- longBits \& DoubleConsts.EXP\_BIT\_MASK |==> longBits | DoubleConsts.EXP\_BIT\_MASK@
@\color{instance-inspecific-input}- longBits \& DoubleConsts.EXP\_BIT\_MASK) >> (DoubleConsts.SIGNIFICAND\_WIDTH - 1) |==> longBits \& DoubleConsts.EXP\_BIT\_MASK) << (DoubleConsts.SIGNIFICAND\_WIDTH - 1)@
@\color{instance-inspecific-input}- 1 |==> 0@
@\color{instance-inspecific-input}- DoubleConsts.SIGNIFICAND\_WIDTH - 1 |==> DoubleConsts.SIGNIFICAND\_WIDTH \% 1@
@\color{instance-inspecific-input}[[Code]]@
@\color{instance-inspecific-input}...@
@\color{instance-inspecific-input}(3 more examples)@
@\color{instance-inspecific-input}...@
@\color{instance-inspecific-input}[[Code]]@
@\color{instance-specific-input}WeightMass mass = lhsDist.get(classVal);@
@\color{instance-inspecific-input}[[Mutations]]@
@\color{model-output}- classVal |==> classVal + 1@
@\color{model-output}- classVal |==> 0@
    \end{lstlisting}
    \vspace{-2ex}
    \caption{Prompt used for mutant generation. We shot the natural language description in \textcolor{NL-description}{purple}, any instance-specific parts of the prompt (e.g., general examples, separators, etc.) in black, inputs that are specific to a concrete instance (i.e., the part that changes when the instance changes) in \textcolor{instance-specific-input}{blue}, and parts that the FSLM model generates in response to that input in \textcolor{model-output}{green}.}
    \label{fig:mutant_prompt}
    \vspace{-2ex}    
\end{figure}


\subsubsection{Prompt}
To generate mutants via an FSLM, we design a prompt that ask the model to modify one code line at a time.
Figure~\ref{fig:mutant_prompt} shows the default prompt for our study.
The prompt contains a brief natural language description of the task to perform, followed by a short list of examples.
To help the model understand the different sections of the prompt, we mark them, e.g., via brackets as in ``[[Code]]''.
Each example consists of the line of code to modify (``[[Code]]'') and a few mutants to generate based on it (``[[Mutations]]'').
Since mutants are small, we ask the model to suggest multiple mutants at once.
Thus, the temperature can be set low for consistent but not as diverse results.
At the end, the prompt provides the code line we wish to mutate, leaving the task of completing it with suitable mutants to the model.
For the example in Figure~\ref{fig:mutant_prompt}, the model suggests a mutant that replaces the expression \code{classVal} passed as parameter in the call \code{lhsDist.get()} with \code{classVal + 1}.


\subsubsection{Post-processing}
Once the model completes the prompt, we post-process the raw completion using simple regular expressions to extract the mutations suggested by the model.
Because an FSLM does not guarantee to produce code that is syntactically or semantically valid, we filter out any suggested mutants that do not compile.
All remaining code snippets are our final set of mutants generated by the FSLM.

\subsubsection{Benchmarks}
\label{sec:mutation benchmarks}
\begin{wraptable}[10]{r}{0.24\textwidth}
  \vspace{-8ex}  
  \small
  \caption{Projects used in our study.}
  \vspace{-2ex}
  \setlength{\tabcolsep}{2pt}    
\label{tab:projects}
\begin{tabular}{@{}lrr@{}}
\toprule
Project       & Version & \# Classes \\ \midrule
Colt          &  1.2.0   &     297              \\
ElasticSearch &  6.1.1   &    2,821               \\
GWT           &  2.5.1   &    3,178                \\
GraphStream   &  1.3   &      233              \\
Guava         &   19.0  &      464               \\
Hibernate     &  5.4.2   &     3,393                \\
JDK           &  8   &    4,240                \\
Commons Math          &  3.6.1   &     918              \\
Weka          &  3.8.0   &    1,648     \\ \bottomrule
\end{tabular}
\end{wraptable}
As code files to mutate, we randomly select 32 classes from the projects listed on Table~\ref{tab:projects}. The smallest class has 19 lines of code, while the longest has 1,346. In total, the classes have 6,747 lines of code.
Across the 32 classes, our instance extractor yields 1,194 instances (lines of code) to generate mutants for.

\subsection{Task 2: Generating Oracles from Natural Language Documentation}
\label{sec:task-two}

As a second code generation task, we consider the problem of generating test oracles from natural language documentation.
This task represents a class of tasks where an FSLM translates natural language to code.
Specifically, we focus on the task of extracting metamorphic test oracles for API methods from Javadoc comments.
A metamorphic test oracle states that two inputs that are in a specific relationship are expected to lead to outputs that are in another known relationship~\cite{Chen1998}.
In the context of testing API methods, this typically means that some API usage is equivalent (or in some other relationship) to some other API usage.
%
As an example of an oracle we aim to generate, consider this excerpt from the Javadoc of the \code{Array.toString} method: ``\emph{The value returned by this method is equal to the value that would be returned by \code{Arrays.asList(a).toString()}, unless \code{a} is null, in which case null is returned.}''.

The equivalence described in this documentation could be specified as an executable test oracle that states that \code{Arrays.toString(a)} yields the same as \code{Arrays.asList(a).toString()} if \code{a != null}.

\subsubsection{Baseline Tool}
Extracting test oracles from natural language documentation is an active area of research~\cite{Blasi2018,Goffi2016,Blasi2021}.
As a baseline to compare an FSLM-based tool against, we use the recently proposed MeMo~\cite{Blasi2021}.
MeMo extracts metamorphic test oracles by first identifying natural language sentences that could contain such oracles using simple heuristics, and then translating those sentences into code.
This translation, which is the most intricate part of MeMo, decomposes the sentence using a dependency parser, and then converts the parsed sentence into code based on a set of hard-coded rules and heuristics.
Because of the inherent imprecision and diversity of natural language, the second step has to cover many edge cases to be effective.
Our study investigates whether an FSLM-based tool could replace or complement this second step of MeMo, i.e., replacing the hard-coded rules by queries to a pre-trained model.

\subsubsection{Instance Extraction}
In the context of this task, we define an instance to be a method whose description probably contains an oracle.
For a fair comparison with the baseline tool, and because extracting such sentences is comparatively simple, we MeMo to identify sentences that likely contain an oracle. We then pass the entire comment containing such a sentence into our prompt, which provides the FSLM with some context.

\begin{figure}[tb]
    \begin{lstlisting}[language=Java,linewidth=8.5cm,basicstyle=\ttfamily\footnotesize,breaklines=true,otherkeywords={describe, it},escapechar={|}]
|\color{instance-inspecific-input}\#\#\# Signature|
|\color{instance-inspecific-input}public static java.lang.String toString(java.lang.Object[] a)|
|\color{instance-inspecific-input}\#\#\# Comment|
|\color{instance-inspecific-input}...\\
The value returned by this method is equal to the value that would\\
be returned by Arrays.asList(a).toString(), unless a is null, in\\
which case "null" is returned.|
|\color{NL-description}\#\#\# Analysis|
|\color{NL-description}This method returns the same thing as the expression\\
Arrays.asList(a).toString(), therefore they are equivalent\\
(at least as long as a is not null).|
|\color{instance-inspecific-input}\#\#\# Equivalence|
|\color{instance-inspecific-input}if (a != null) {{ toString(a) <-> Arrays.asList(a).toString() }};|
|\color{instance-inspecific-input}...|
|\color{instance-inspecific-input}(3 more examples)|
|\color{instance-inspecific-input}...|
|\color{instance-inspecific-input}\#\#\# Signature|
|\color{instance-specific-input}public double norm2(cern.colt.matrix.DoubleMatrix1D x)|
|\color{instance-inspecific-input}\#\#\# Comment|
|\color{instance-specific-input}Returns the two-norm (aka euclidean norm) of vector x;\\
equivalent to mult(x,x).|
|\color{NL-description}\#\#\# Analysis|
|\color{model-output}This method is equivalent to the expression mult(x,x).|
|\color{instance-inspecific-input}\#\#\# Equivalence|
|\color{model-output}norm2(x) <-> mult(x,x);|
    \end{lstlisting}
    \vspace{-2ex}
    \caption{The prompt used for oracle extraction.}
    \label{fig:oracle_prompt}
    \vspace{-3ex}    
\end{figure}

\subsubsection{Prompt}
\label{sec:task2 prompt}

We design the prompt to be a short list of examples of what the model is supposed to achieve, as shown in Figure~\ref{fig:oracle_prompt}.
Each example consists of four parts: (1) The signature of the method for which to generate an oracle (``\#\#\# Signature''), (2) the natural language description of the method's behavior, as extracted from the available Javadoc (``\#\#\# Comment''), (3) A small section of natural language explanation about how the equivalence manifests itself in the example (``\#\#\# Analysis'') (This part is motivated by the observation that by letting the model explain its reasoning before generating the result itself may increase its effectiveness~\cite{wei2022chain}.), and (4) the Java code of the metamorphic oracle, which consists of a conditional followed by two expressions separated by the symbol \code{<->}, denoting ``equivalent to'' (``\#\#\# Equivalence'').
After providing a small number of such examples (four by default), we provide the signature and comment of the instance we are interested in, and then let the model complete the prompt by providing an analysis and the oracle.
For this task, the temperature is set to zero, as we observe the model to produce too imprecise predictions otherwise.

\subsubsection{Post-processing}
Given the raw completion produced by the model in response to our prompt, we extract the generated test oracle.
The extraction is based on simple regular expressions, e.g., anchored around the special \code{<->} symbol.
Next, we check whether the predicted condition (if any) and the code snippets compile properly.
Finally, the approach expands names of classes, e.g., \code{Math} to \code{java.lang.Math}, using JavaSymbolSolver.\footnote{\url{http://javaparser.org/}}

\subsubsection{Benchmarks}
\label{sec:benchmarks oracles}

To measure the effectiveness of the FSLM-based tool, we use a ground truth dataset available from MeMo's artifacts~\cite{Blasi2021}.
The dataset is based on 5,000 methods from nine open-source Java projects, from which 299 metamorphic test oracles have been manually extracted.
The oracles are diverse and vary in length:
The natural language descriptions range between 3 and 500 words, with a mean of 44.3.
The code of the oracles ranges between 3 and 81 tokens, with a mean of 21.6.



\subsection{Task 3: Test Case Generation}
\label{sec:task-three}

As the third code generation task, we consider the problem of
generating unit tests. This task represents a class of tasks where the FSLM generates a  method, i.e., a larger portion of code compared with the previous examples. Test case generation is a labor-intensive task
in software testing~\cite{ANAND20131978}, and
several techniques have been proposed to automate unit test case generation~\cite{Serra2019}.


\subsubsection{Baseline Tool}
There are many automatic test case generation tools available. Randoop~\cite{Pacheco2007} and EvoSuite~\cite{Fraser2011a} are popular representatives of such tools. We use Randoop in our study.
To generate test cases with Randoop, for each method under test, we invoke its main class \code{randoop.main.Main} passing the \code{gentests} command and the \code{--methodlist=filename.txt} and \code{--generated-limit=100} arguments. The file \code{filename.txt} contains the method under test, as well as helper methods it depends on. We select helper methods with a minimum amount of dependencies to include. The \code{generated-limit} argument defines the maximum number of test method candidates generated internally.
For a fair comparison, we let Randoop and the FSLM generate the same number (100) of test cases per method under test.


\subsubsection{Instance Extraction}
For unit test case generation, we consider an instance to be a method under test.
That is, the instance extractor takes a Java class as its input, produces a list of public methods, and randomly selects a method from the list to be tested.

\subsubsection{Prompt}
\label{sec:test-generation-prompt-section}
Figure~\ref{fig:test-generation-prompt} shows an example of the (default) prompt that we use for unit test case generation.
The prompt starts with a brief natural language description of the task.
Next, we provide one example of the task.
The reason for showing only one example is that state-of-the-art FSLMs only support a bounded prompt size.
The example consists of three parts: (1) a list of helper methods to assist in the creation of values. (``Helper constructors and methods:''), (2) the method under test itself, and (3) a test case that invokes the method under test.
After the example, the instance, consisting of the code of the method (as explained on Section 3.4.2 ``Instance Extraction'') is provided, leaving the task of generating a test case to the model. Since the prompt contains only a single example, selecting this example potentially has a large impact on the generated test.
Section~\ref{sec:answer rq2} compares different strategies for selecting the example, e.g., selecting another method under test from the same class and selecting another method under test at random.
Because each query yields only one test case, we make multiple queries while varying the temperature parameter from 0.0 to 0.9, in steps of 0.1.
For each temperature, we make \numberOfQueriesPerTemperature{} queries.
This way, the model predicts a total of \numberOfCandidateTests{} test case candidates.

\begin{figure}[tb]
    \begin{lstlisting}[language=Java,linewidth=8.5cm,basicstyle=\ttfamily\footnotesize,breaklines=true,escapechar={|}]
|\color{NL-description}Suggest a test for a method with the DoubleArrayList quantiles(DoubleArrayList percentages) signature.|
|\color{instance-inspecific-input}Helper constructors and methods:|
  |\color{instance-specific-input}DynamicBin1D()|
  |\color{instance-specific-input}DoubleArrayList()|
|\color{instance-inspecific-input}Method:| |\color{instance-specific-input}public synchronized double max() \{|
  |\color{instance-specific-input}if (! isIncrementalStatValid) updateIncrementalStats();|
  |\color{instance-specific-input}return this.max;|
|\color{instance-specific-input}\}|
|\color{instance-specific-input}Test: public static void testMax() \{|
  |\color{instance-specific-input}double[] temp = new double[2];|
  |\color{instance-specific-input}temp[0] = 8.9;|
  |\color{instance-specific-input}temp[1] = 1;|
  |\color{instance-specific-input}DenseDoubleMatrix1D d1Double = new DenseDoubleMatrix1D(temp);|
  |\color{instance-specific-input}hep.aida.bin.DynamicBin1D d1ynamicBin = cern.colt.matrix.doublealgo.Statistic.bin(d1Double);|
  |\color{instance-specific-input}double max = d1ynamicBin.max();|
  |\color{instance-specific-input}System.out.println("max="+ max);|
|\color{instance-specific-input}\}|
|\color{instance-inspecific-input}---|
|\color{instance-inspecific-input}Method:||\color{instance-specific-input} public DoubleArrayList quantiles(DoubleArrayList percentages)\\ \{|
  |\color{instance-specific-input}return Descriptive.quantiles(sortedElements\_unsafe(),percentages);|
|\color{instance-specific-input}\}|
|\color{instance-inspecific-input}Test:||\color{model-output}public static void testQuantiles() \{|
  |\color{model-output}double[] temp = new double[2];|
  |\color{model-output}temp[0] = 8.9;|
  |\color{model-output}temp[1] = 1;|
  |\color{model-output}DenseDoubleMatrix1D d1Double = new DenseDoubleMatrix1D(temp);|
  |\color{model-output}hep.aida.bin.DynamicBin1D d1ynamicBin = cern.colt.matrix.doublealgo.Statistic.bin(d1Double);|
  |\color{model-output}DoubleArrayList quantiles = d1ynamicBin.quantiles(new DoubleArrayList(new double[] {0.5,0.75}));|
  |\color{model-output}System.out.println("quantiles="+ quantiles);|
|\color{model-output}\}|
    \end{lstlisting}   
    \vspace{-2ex}
    \caption{Test generation prompt for the method \texttt{DoubleArrayList quantiles(DoubleArrayList percentages)}, declared in the class \texttt{DynamicBin1D} from project Colt.}
    \label{fig:test-generation-prompt}
\end{figure}

\subsubsection{Post-processing}
To post-process the raw completions of the model, we inject each test case candidate into a template of a test class, which contains the necessary scaffolding to yield an executable test case. 
Similar to the previous tasks, we discard candidates that do not compile. 
We also remove any duplicates that may result from querying the model multiple times.

\subsubsection{Benchmarks}
\label{sec:benchmarks test gen}

As methods under test we use the \numberOfMUT{} methods that Table~\ref{tab:test-coverage-analysis-davinci-and-randoop-tests} shows.
We select them by randomly identifying \numberOfTestsPerMUT{} public methods from each of the \numberOfProjects{} projects in Table~\ref{tab:projects}.


\section{Results}

This section presents answers to the three research questions posed in
Section~\ref{sec:rqs}. Section~\ref{sec:discussion} discusses the
results and their broader impact.

\subsection{RQ1: Accuracy}

This section presents results on the accuracy of FSLM-based code generation compared to traditionally built tools.


\subsubsection{Code Mutation}

\begin{table*}
    \centering
    \setlength{\tabcolsep}{3pt}
    \caption{Mutants generated by our FSLM-based tool and by Major~\cite{major}.}
    \setlength{\tabcolsep}{7pt}
\begin{tabular}{@{}lrrr|rrrr@{}}
\toprule
& \multicolumn{3}{c}{Generated mutants} & \multicolumn{4}{c}{Kind of transformation} \\
\cmidrule{2-4}
\cmidrule{5-8}
& Total & Overlap w/ Major & Compilable & Delete statement & Replace operator & Replace value & Other \\
\midrule
               FSLM &    \textbf{2,721} &   18.4\%  &   62.5\%  & 31.8\% &                   9.0\% &               34.7\% &            24.5\%  \\
FSLM (NL-only) & 0 & - & - &- &- &- &- \\
FSLM (Ex-only) &    2,645 &   17.7\%  &   57.6\%  & 35.2\% &                   7.0\% &               36.3\% &            21.6\%       \\
FSLM (Bad-ex) &    2,595 &   19.0\%  &   64.5\%  & 29.4\% &                   9.1\% &               37.1\% &            24.4\%      \\
FSLM (Small model) &    2,487 &   15.4\%  &   53.8\%  & 30.0\% &                   4.2\% &               33.1\% &            32.8\%       \\
              Major &    \textbf{2,810} &  100.0\%  &  100.0\%  &  4.9\% &                  48.6\% &               35.8\% &             0.0\%      \\
\bottomrule
\end{tabular}

    \label{tab:mutants_comparison_table}
\end{table*}

Table~\ref{tab:mutants_comparison_table} summarizes our results for the code mutation task.
Given the 1,194 instances extracted from 32 classes (Section~\ref{sec:mutation benchmarks}), our FSLM-based tool generates a total of 2,721 mutants, whereas the baseline Major tool generates 2,810 mutants.
Because the model does not guarantee to generate valid code, only 62.5\% of the FSLM-generated mutants are compilable, giving a total of 1,701 usable mutants.
On average, our tool changes 3.97 tokens of the original code, which roughly equals the 4.28 tokens changed by Major.
Besides the raw amount of mutants generated, it is also important to understand whether the generated mutants are useful.
We address this question both quantitatively and qualitatively.
As a quantitative answer, we compute how many of the FSLM-generated mutants exactly match one of the Major-generated mutants.
We observe an overlap of around 18\% of the FSLM-generated mutants.
Under the assumption that Major-generated mutants are useful, this means that at least 18\% of the FSLM-generated mutants are also useful.

As a qualitative answer, we manually inspect a random sample of 30 of the compilable mutants our tool generates.
For each sampled mutant, we carefully inspect the code and determine whether the mutation changes the runtime behavior of the code, as opposed to being an equivalent mutant that preserves the semantics of the original code.
The inspection shows that 90\% of the mutants certainly change the behavior, whereas the remaining 10\% either preserve the semantics or we could not clearly determine its effects.

To better understand the mutants generated by the model and Major, we automatically classify them based on the kind of code transformation.
We distinguish four classes, as shown in the four right-most columns of Table~\ref{tab:mutants_comparison_table}: 
(i) deleting a statement, (ii) replacing one operator with another, (iii) replacing one value with another, and (iv) some other transformation.
The table shows that the distribution of mutants that the FSLM and Major generate clearly differ: While Major mostly replaces operators and values, the model generates a more diverse set of mutants, suggesting that the two tools complement each other.

Finally, we manually study another random sample of 30 mutants produced by each tool to get qualitative insights into the differences between the two tools.
We make two interesting observations:
\begin{itemize}
\item The FSLM model generates mutants that Major cannot generate based on its built-in mutation operators~\cite{major}. For example, these FSLM-generated mutants include adding a constant to an integer (e.g., turning \code{nanos} into \code{nanos + 1}) and changing methods to semantically similar ones (e.g., turning \code{Math.min} into \code{Math.max}).
\item A relatively large amount of the FSLM-generated mutants (7/30=23\%) replace an expression with \code{null}. While this yields mutant that change the semantics, the high amount is still surprising.
\end{itemize}

Overall, these results show that our FSLM-based tool, while not generating exactly the same mutants as an existing tool, nonetheless creates a large number of useful mutants with minimal effort.

\subsubsection{Generating Oracles from Natural Language Documentation}

This section compares the accuracy of (metamorphic) test oracle generators, namely, the state-of-the-art MeMo~\cite{Blasi2021} and its FSLM-based counterpart.
To measure accuracy, we compare all generated oracles against a ground truth consisting of 299 test oracles that we wish to extract from the documentation of methods from the projects listed on Table~\ref{tab:projects}. Specifically, we measure precision ($Pr$) and recall ($Re$) as follows:

\noindent
\begin{tabular}{@{}ll}
$Pr = \frac{\text{\# of correctly generated oracles}}{\text{\# of all generated oracles}}$ &
$Re = \frac{\text{\# of correctly generated oracles}}{\text{\# of all ground truth oracles}}$  
\end{tabular}

\begin{table}
    \centering
    \setlength{\tabcolsep}{7pt}
    \caption{Effectiveness of test oracle generation.}
    \setlength{\tabcolsep}{2.5pt}
\begin{tabular}{@{}lrrr|rrr@{}}
\toprule
\multicolumn{1}{c}{\multirow{2.5}{*}{Project}} & \multicolumn{3}{c}{FSLM} & \multicolumn{3}{c}{MeMo~\cite{Blasi2021}} \\
\cmidrule{2-4} \cmidrule{5-7}
& Precision & Recall & F1 & Precision & Recall & F1 \\
\midrule
           JDK &      0.84 &   0.54 &     0.66 &      0.50 &   0.55 &     0.52 \\
          Colt &      0.65 &   0.42 &     0.51 &      0.61 &   0.42 &     0.50 \\
         Guava &      0.88 &   0.52 &     0.65 &      0.83 &   0.67 &     0.74 \\
           GWT &      0.79 &   0.27 &     0.40 &      0.65 &   0.27 &     0.38 \\
   Graphstream &      1.00 &   0.70 &     0.82 &      1.00 &   0.70 &     0.82 \\
Apache commons &      0.79 &   0.58 &     0.67 &      0.66 &   0.82 &     0.73 \\
     Hibernate &      0.00 &   0.00 &     0.00 &      0.00 &   0.00 &     0.00 \\
 Elasticsearch &      0.00 &   0.00 &     0.00 &      0.00 &   0.00 &     0.00 \\
          Weka &      0.50 &   0.50 &     0.50 &      0.43 &   0.50 &     0.46 \\
          \midrule
           Total &      0.82 &   0.47 &   0.60 &      0.64 &   0.54 &     0.59 \\
\bottomrule
\end{tabular}

    \label{tab:oracle_memo_comparison_table}
\end{table}

In addition, we report the F1-score, defined as the harmonic mean of precision and recall. Table~\ref{tab:oracle_memo_comparison_table} shows the results for each of the studied libraries.
Across all projects, the FSLM-based oracle generator achieves an F1-score of 0.60, which slightly outperforms MeMo's F1-score of 0.59.
Comparing precision and recall shows that the model tends to generate oracles much more precisely, with a precision of 0.82 instead of MeMo's precision of 0.64.

\sloppy
To understand the strengths and weakness of the two approaches, we manually study some of the oracles.
On the one hand, we inspect those oracles that the model predicts correctly while MeMo misses them, which are nine oracles in total.
Three of the nine oracles are cases where there exist multiple oracles for a single method, and the model discovers one, whereas MeMo discovers the other.
This highlights a limitation of our prompt design, which enables the model to predict only one oracle per method.
One could remedy this limitation by prompting for a list of oracles, similar to the prompt for code mutations, or by querying the model multiple times with a higher temperature, similar to what we do for test generation.
The remaining six oracles are all related to MeMo incorrectly capturing longer or nested pieces of code.
For example, the documentation ``Calling getSource() is equivalent to calling getSnapshot().getSource()'' is translated by MeMo to an equivalence between \code{getSource()} and \code{getSnapshot()}, which is incorrect.
In contrast, the model correctly predicts the equivalence between \code{getSource()} and \code{getSnapshot().getSource()}.

On the other hand, we also inspect the six instances where the model misses an oracle that MeMo can predict. For two of these oracles, the model ``invents'' code seemingly out of thin air.
For example, the documentation ``This is equivalent to, but not necessarily implemented as, !(Float.isInfinite(value) |~| Float.isNaN(value)).'' leads to an incorrect prediction of the model saying that \code{com.google.common.primitives.Float.isFinite} and \code{java.lang.Float.isFinite} are equivalent.

Overall, the FSLM-based oracle generator achieves results that are on par, and even slightly better, than those of a state-of-the-art tool based on a set hard-coded rules and heuristics.

\subsubsection{Test Case Generation}

\begin{table*}[]
\caption{\label{tab:test-coverage-analysis-davinci-and-randoop-tests} Analysis of the test cases generated by our FSLM-based test generator and Randoop for the considered methods. The table presents (1) the number of compilable test (CT) cases; (2) the average test size (TS) of the generated tests and (3) the line coverage (LC) achieved by them.}
\setlength{\tabcolsep}{6pt}
\begin{tabular}{@{}llrrrrrrr@{}}
\toprule
\multicolumn{1}{c}{\multirow{3}{*}{Project}} & \multicolumn{1}{c}{\multirow{3}{*}{Method}}                  & \multicolumn{3}{c}{\multirow{2}{*}{FSLM}}                                  & \multicolumn{3}{c}{\multirow{2}{*}{Randoop}}              & \multicolumn{1}{c}{\multirow{2}{*}{Combined}} \\
\multicolumn{1}{c}{}                         & \multicolumn{1}{c}{}                                         & \multicolumn{3}{c}{}                                                           & \multicolumn{3}{c}{}                                                           & \multicolumn{1}{c}{}                                        \\ \cline{3-9} 
\multicolumn{1}{c}{}                         & \multicolumn{1}{c}{}                                         & \multicolumn{1}{c}{\# CT} & \multicolumn{1}{l}{TS} & \multicolumn{1}{c|}{LC}   & \multicolumn{1}{c}{\# CT} & \multicolumn{1}{l}{TS} & \multicolumn{1}{c|}{LC}   & \multicolumn{1}{c}{LC}                                      \\ \midrule
Colt                                         & DoubleArrayList quantiles(DoubleArrayList percentages)       & 12                        & 11                     & \multicolumn{1}{r|}{29\%} & 29                        & 15                     & \multicolumn{1}{r|}{26\%} & 39\%                                                        \\
Colt                                         & double moment(int k, double c)                               & 7                         & 13                     & \multicolumn{1}{r|}{17\%} & 49                        & 16                     & \multicolumn{1}{r|}{15\%} & 19\%                                                        \\
ElasticSearch                                & String parent()                                              & 4                         & 4                      & \multicolumn{1}{r|}{8\%}  & 1                         & 202                    & \multicolumn{1}{r|}{5\%}  & 8\%                                                         \\
ElasticSearch                                & IndexRequest source(Map source, XContentType contentType)    & 8                         & 10                     & \multicolumn{1}{r|}{13\%} & 66                        & 35                     & \multicolumn{1}{r|}{5\%}  & 13\%                                                        \\
GWT                                          & boolean isClient()                                           & 3                         & 7                      & \multicolumn{1}{r|}{6\%}  & 1                         & 7                      & \multicolumn{1}{r|}{6\%}  & 6\%                                                         \\
GWT                                          & UncaughtExceptionHandler getUncaughtExceptionHandler()       & 0                         & 0                      & \multicolumn{1}{r|}{0\%}  & 1                         & 7                      & \multicolumn{1}{r|}{6\%}  & 6\%                                                         \\
Graphstream                                  & boolean contains(Edge edge)                                  & 1                         & 12                     & \multicolumn{1}{r|}{12\%} & 49                        & 85                     & \multicolumn{1}{r|}{9\%}  & 13\%                                                        \\
Graphstream                                  & boolean equals(Path p)                                       & 42                        & 17                     & \multicolumn{1}{r|}{31\%} & 44                        & 112                    & \multicolumn{1}{r|}{11\%} & 31\%                                                        \\
Guava                                        & HashCode fromLong(long hash)                                 & 6                         & 8                      & \multicolumn{1}{r|}{12\%} & 28                        & 12                     & \multicolumn{1}{r|}{7\%}  & 12\%                                                        \\
Guava                                        & int writeBytesTo(byte{[}{]} dest, int offset, int maxLength) & 37                        & 10                     & \multicolumn{1}{r|}{32\%} & 50                        & 23                     & \multicolumn{1}{r|}{17\%} & 34\%                                                        \\
Hibernate                                    & Short toShort(Boolean value)                                 & 3                         & 4                      & \multicolumn{1}{r|}{24\%} & 3                         & 6                      & \multicolumn{1}{r|}{20\%} & 24\%                                                        \\
Hibernate                                    & Boolean fromString(String string)                            & 17                        & 7                      & \multicolumn{1}{r|}{22\%} & 35                        & 19                     & \multicolumn{1}{r|}{22\%} & 22\%                                                        \\
JDK                                          & Object remove(int index)                                     & 13                        & 15                     & \multicolumn{1}{r|}{5\%}  & 70                        & 19                     & \multicolumn{1}{r|}{3\%}  & 5\%                                                         \\
JDK                                          & boolean contains(Object o)                                   & 7                         & 9                      & \multicolumn{1}{r|}{3\%}  & 44                        & 26                     & \multicolumn{1}{r|}{1\%}  & 3\%                                                         \\
Math                                         & Vector1D scalarMultiply(double a)                            & 28                        & 7                      & \multicolumn{1}{r|}{30\%} & 58                        & 10                     & \multicolumn{1}{r|}{24\%} & 32\%                                                        \\
Math                                         & double distanceSq(Vector1D p1, Vector1D p2)                  & 9                         & 6                      & \multicolumn{1}{r|}{23\%} & 51                        & 17                     & \multicolumn{1}{r|}{18\%} & 24\%                                                        \\
Weka                                         & AlgVector add(AlgVector other)                               & 5                         & 10                     & \multicolumn{1}{r|}{28\%} & 47                        & 35                     & \multicolumn{1}{r|}{17\%} & 28\%                                                        \\
Weka                                         & Instance getAsInstance(Instances model, Random random)       & 0                         & 0                      & \multicolumn{1}{r|}{0\%}  & 56                        & 15                     & \multicolumn{1}{r|}{8\%}  & 8\%                                                         \\ \midrule
Total                                        &                                                              & 202                       & 11                     & \multicolumn{1}{r|}{14\%} & 682                       & 31                     & \multicolumn{1}{r|}{10\%} & 16\%                                                        \\ \bottomrule
\end{tabular}
\end{table*}

Table~\ref{tab:test-coverage-analysis-davinci-and-randoop-tests} summarizes the results of generating test cases with our FSLM-based approach and with Randoop~\cite{randoop} on \numberOfMUT{} methods.
The table reports the amount of compilable tests (column ``CT''), the average size of tests in number of lines of code (column ``TS''), and the line coverage that the tests achieve (column ``LC'').
We measure coverage using JaCoCo~\cite{jacoco}. We notice from these results that, overall, the model achieves higher code coverage than Randoop (14\% vs. 10\%).
This result is particularly remarkable as Randoop generates more than three times the number of tests the model generates (202 vs. 682 tests). Moreover, on average, the size of the tests generated by the model are much smaller than the tests generated by Randoop (11 vs. 31 lines).

On a closer analysis of the tests generated by each approach, for each of the \numberOfMUT{} methods, we can see that Randoop successfully generates tests for all \numberOfMUT{} methods under test.
In contrast, the model successfully generates tests for only 16 of them. More specifically, (i)~for 14 methods, the tests generated by the model achieve higher coverage than the tests generated by Randoop; (ii)~for two methods, the tests generated by both approaches achieve the same coverage; (iii)~for two methods, the tests generated by Randoop achieve higher coverage than the tests generated by the model. These are exactly the two methods for which the model fails to generate any compilable tests.

These results provide initial evidence indicating that FSLM-based tools can outperform state-of-the-art test generation tools.
We also calculate the coverage achieved by combining the tests generated by both approaches. The results can be seen in the last column of Table~\ref{tab:test-coverage-analysis-davinci-and-randoop-tests}. Interestingly, the coverage achieved by the combination of the tests (16\%) is superior to the coverage achieved by the tests of each approach individually. As an example, the coverage achieved by the combination of the tests is considerably higher when considering the \code{quantiles} method of the Colt project. In this case, individually, the tests generated by the model achieve 29\% line coverage and the tests generated by Randoop achieve 26\% line coverage. Combined, the tests generated by both approaches achieve 39\% line coverage.

\begin{center}
\fbox{
\begin{minipage}{3 in}
{Summary of \textbf{RQ1}}: {\it FSLM-based tools perform surprisingly well across the different tasks, being on par or complimentary, and for test generation even better, than the handcrafted tools we compare against.}
\end{minipage}
}
\end{center}

\subsection{RQ2: Impact of Prompt}
\label{sec:answer rq2}


By default, our prompts contain both natural language task descriptions and input-output examples. This section reports on the impact of using different prompt variants. For each of the tasks, we consider the following prompt variants: Only natural language description (NL-only); Only input-output examples (Ex-only); Poorly chosen examples (Bad-ex).

\subsubsection{Code Mutation}
For mutant generation, NL-only means the prompt includes only the natural language text at the top of Figure~\ref{fig:mutant_prompt}, Ex-only means we keep everything but the NL description, and Bad-ex means we include additional examples where our FSLM-based tool should not generate mutants. For example, we add an import statement as an example, but leave the mutants section empty.
The idea is to test how robust the model is to adversarial or poorly chosen examples.

The middle rows in Table~\ref{tab:mutants_comparison_table} show the results obtained with these variants.
Running NL-only does not produce promising results since it is missing the guiding output format from the examples. We attempt to "fix" the prompt by including more detailed descriptions on how to format the output (i.e. we add "Return the result in the format original |==> replacement as part of a list numbered using '-'." to the prompt), but the output format remains inconsistent, giving no results. This means examples play a large part in solving this task using an FSLM.
Looking at the results for Ex-only reveals that less generated mutants compile, with a margin of 5\%. This is interesting as the textual description is only a single sentence in this task and shows an easy way to improve performance over using a prompt without it.
Moreover, we observe the following behavior for the Bad-ex variant of the prompt. The overlap with Major and percentage of mutants that compile are actually slightly higher than for our default approach. This is surprising in that a deliberate attempt to worsen the predictions instead slightly improves the results.
%

\subsubsection{Generating Oracles from Natural Language Documentation}
For oracle generation, NL-only means we only add a natural language description of the task and some information about formatting (e.g., "Extract equivalent pieces of code from the following comments. Format the result as Code snippet A <-> Code snippet B."). For Ex-only we remove the part of the prompt that describes the task in NL (see purple text on Figure~\ref{fig:oracle_prompt}). This is different from the style employed for mutant generation though, as in the oracle extraction prompt the natural language description is part of each example and not just a general task description. For Bad-ex, we once again add examples designed to throw off the FSLM by including examples that the model should not generate anything for. For example, we add a method with comment ``Returns the largest value of the given array.'' and leave the oracle section empty.

\begin{figure}
    \centering
    \hspace*{-2.4cm}\includegraphics[width=13cm]{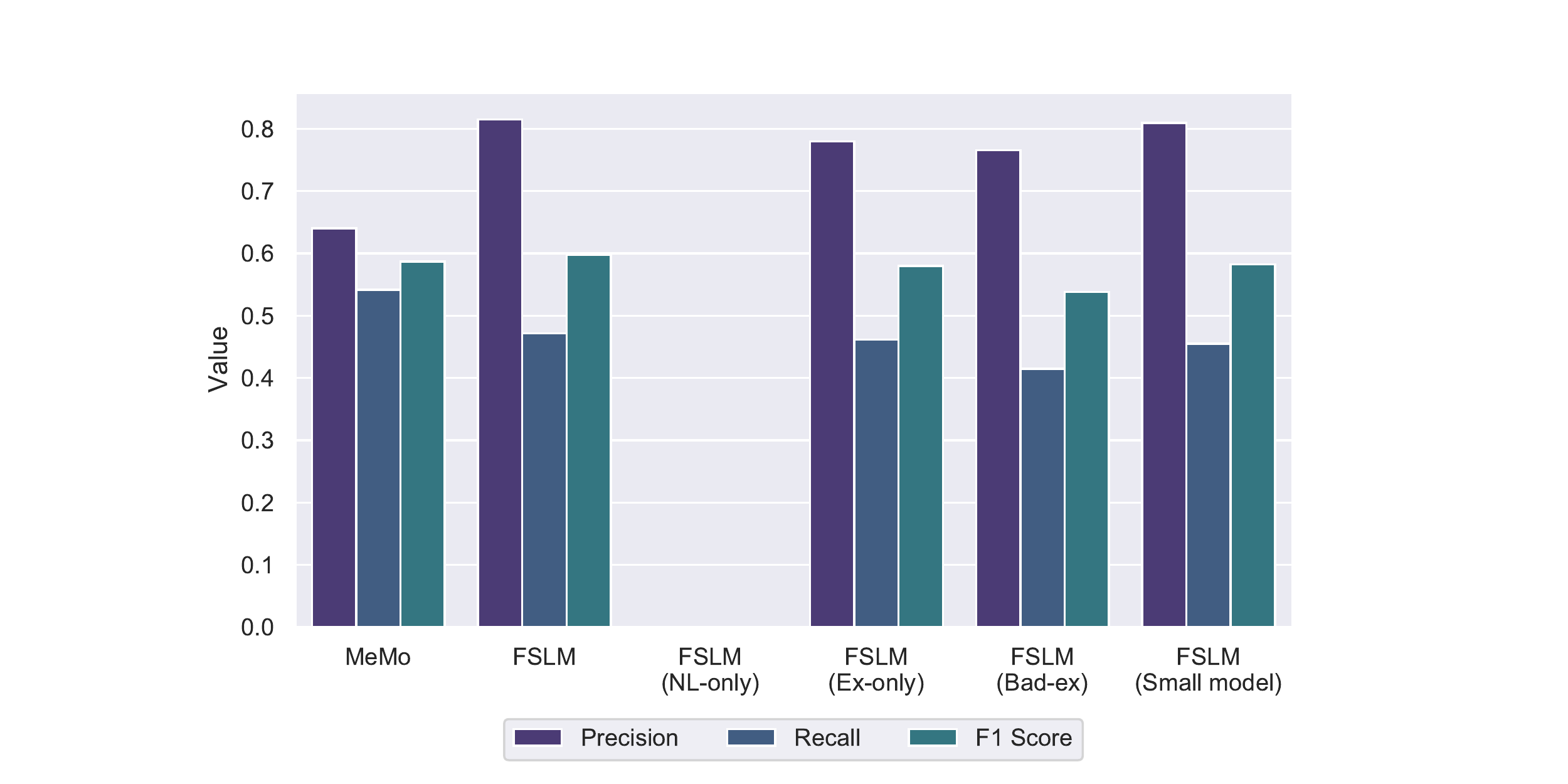}
    \vspace{-4ex}    
    \caption{Oracle generation results. Comparison of prompt variations.}
    \label{fig:oracles_comparison_plot}
    \vspace{-3ex}
\end{figure}

Figure~\ref{fig:oracles_comparison_plot} shows results of the FSLM for the oracle generation task, when using different prompt variations. The accuracy is not significantly affected by the different styles of prompt used, except for NL-only. As for mutant generation, NL-only yields incorrectly formatted responses, giving no usable results. Again, examples appear necessary to be able to successfully use FSLMs for this task.
Considering the prompt variant where we remove NL description, Ex-only, we observe that the difference in performance is negligible compared to the default prompt, indicating that the helping text is not as important as it was for mutation generation.
Considering the prompt variant Bad-ex, we observe that the use of bad examples performs worse compared to other types of prompts. This indicates that the quality of examples for this task is more important than for mutant generation.

A likely explanation for this discrepancy across the tasks is that the way natural language descriptions are used in the second task differs from how it is used in the other two tasks (Section~\ref{sec:task2 prompt}). Consequently, to more uniformly compare the tasks, we also run an experiment with a prompt where the natural language description is in the form of a task description. This prompt yields an F1 score of 0.54, i.e., substantially worse than our default prompt. These results suggest that the quality of the examples are relatively more important than the NL description for this task.

\subsubsection{Test Case Generation}

\begin{table}[t]
  \caption{\label{tab:test-coverage-for-different-prompts} Line coverage achieved by the tests generated by our FSLM-based test generator with different prompts and a smaller model.}
  \vspace{-1ex}
\setlength{\tabcolsep}{17pt}
\begin{tabular}{@{}lr@{}}
\toprule
Variant of the approach                       & \multicolumn{1}{c}{Line coverage} \\ \midrule
FSLM                   & 14\%                              \\
FSLM w/o example (NL-only)      & 12\%                              \\
FSLM w/o NL descr. (Ex-only)    & 9\%                               \\
FSLM w/ random example (Bad-ex) & 8\%                               \\
FSLM w/ small model & 10\% \\ \bottomrule
\end{tabular}
  \vspace{-2ex}
\end{table}

In this task, the method under test and a list of helper constructors and methods is always provided to the prompt. Therefore, for NL-only we remove the input-output example, for Ex-only we remove the natural language description, and for Bad-ex we provide a different input-output example, which we randomly select from a different project.

Table~\ref{tab:test-coverage-for-different-prompts} reports the results of these experiments. Overall, we can see that, regarding line coverage, the default prompt achieves a higher line coverage of 14\%, followed by variation (NL-only) with 12\% coverage, then variation (Ex-only) with 12\% coverage, and finally variation (Bad-ex) with only 8\% coverage. These results indicate that a natural language description can be even more important than an input-output example for test generation (12\% vs. 9\%). Moreover, an input-output example more related to the method under test, i.e., from the same class in our case, can add more value than a random example unrelated to the method under test (14\% vs. 8\%).

\begin{center}
\fbox{
\begin{minipage}{3 in}
{Summary of \textbf{RQ2}}: {\it Adding a brief natural language description is an easy way to help (or at least not hurt) the FSLM-based tools. Furthermore, we find that providing suitable examples is crucial for the model to make effective predictions.}
\end{minipage}
}
\end{center}

\subsection{RQ3: Impact of Model Size}

Training larger language models on more data often results in performance improvements for downstream tasks~\cite{Brown2020}.
By default, we use the ``Davinci'' model of Codex, which currently is the largest model offered via the OpenAI API.
Since larger models come with a hefty computational price~\cite{DBLP:journals/cacm/HellendoornS22}, we also measure the impact of using a smaller model.
To this end, we repeat our experiments with the ``Cushman'' model of Codex, which is a derivative of a small model trained by Chen et al.~\cite{Chen2021}.


\subsubsection{Code Mutation}
The ``FSLM w/ small model'' row of Table~\ref{tab:mutants_comparison_table} shows the impact of using a smaller model on code mutation.
Several metrics of success clearly drop, e.g., the total number of generated mutants (from 2,721 to 2,487) and the number of mutants that compile (from 62.5\% to 52.8\%).
These results show that using a larger model is beneficial for this task.

\subsubsection{Generating Oracles from Natural Language Documentation}
When running the test oracle generation using a smaller model, we discover, surprisingly, that the results we obtain are nearly identical to the larger model with an F1 score of 0.58 (as compared to 0.6).
Hence, it seems some tasks can be handled by smaller models almost as well as with larger models.

\subsubsection{Test Case Generation}

For test case generation, we observe a significant drop in effectiveness when using the smaller model (Table~\ref{tab:test-coverage-for-different-prompts}).
The line coverage drops to 10\%, i.e., four percent points less than with the larger model and about the same as with tests generated by Randoop.

\begin{center}
\fbox{
\begin{minipage}{3 in}
{Summary of \textbf{RQ3}}: {\it Increasing the model size improves effectiveness, or at least does not negatively affect it, for all three code generation tasks. For one of the three tasks (oracle generation), the effect is small though. Given the computational cost of large models, carefully selecting them for each task is recommended}.
\end{minipage}
}
\end{center}

\section{Discussion}
\label{sec:discussion}

\paragraph{Prompt design} Designing ``good'' prompts is central to the creation of FSLM-based tools. When answering RQ2, we observe that examples are very important in prompt design and that natural language descriptions are often helpful. There are, however, questions that remain to be evaluated, including (i)~how to mine good examples to create prompts, (ii)~whether or not alternating through examples is useful when the user queries the model multiple times, and (iii) how sensible the prompts are to the data format. 

\paragraph{Model size}
Training large-scale models of code may easily cost hundreds, or even
millions, of
dollars~\cite{DBLP:journals/cacm/HellendoornS22}. Additionally, these large-scale models are hard to use due to their sheer size, or not being open to the public in the first place. For our work, we find these models to be effective, but obtaining the same results with an improved smaller, open model would make the tools more accessible in the long run.

\paragraph{Integrating FSLM-based and traditional tools} The conjunction of low effort to create new code generation tools and the promising results we obtain indicate that integrating FSLM-based tools with existing tools can be helpful. For example, the results for the oracle generation task (Table~\ref{tab:oracle_memo_comparison_table}) show different precision-recall tradeoffs of the two tools. Blending FSLM-based and traditional techniques seems a promising direction to explore in the future.

\paragraph{Threats to Validity}
We do not compare our results across different models (except by size), potentially limiting the generalizability of our findings.
%
While we try to evaluate on a diverse set of tasks, there are obviously many more code generation tasks not studied here.
The fact that the FSLM-based approach is able to provide promising results on the first three tasks we study, gives at least some indication about the potential for other tasks.
Finally, we only evaluated Java-based tools, i.e., our results might not generalize beyond this language. Prior research shows that large-scale models perform well across many differing languages~\cite{Chen2021}.

\section{Related Work}

\paragraph{Studies of neural models of code}

As neural models of code become more popular for a diverse set of tasks, many, similar to us, have begun investigating the details of these models. This comes in multiple forms, such as evaluating a series of similar models~\cite{Xu2022} or models with the same architecture but differing size~\cite{Chen2021}.
Another approach is to apply a model of code to multiple downstream tasks and compare its performance, e.g., by fine-tuning a transformer model to perform tasks similar to the ones we explore in our research~\cite{Mastropaolo2021}.
What sets this paper apart is that (1) we investigate few-shot learning, requiring less training data as compared to fine-tuning, (2) we compare against commonly used traditional tools, while others compare neural approaches against each other, and (3) we target a different set of tasks.

\paragraph{Language models in software engineering}

\citet{Degiovanni2022} use a pre-trained language model for mutation testing by masking one token at a time and asking the model to predict an alternative, which is then considered a mutation. Instead, we study using a generative model for end-to-end mutant generation, which often changes multiple tokens at a time.
Several papers~\cite{Austin2021,Chen2021} study language model-based code generation from short natural language descriptions. In contrast to our work, there offer no comparison to traditional tools and focus only on this single task.
\citet{Jain2022} use generative language models for program synthesis given a natural description of the desired functionality and some code examples that are likely similar to the expected code.
They propose a ``context bank'' of examples to provide in the prompt, which is an idea one could also adapt for our tasks.

\paragraph{Generative language models in general}
Since the introduction of Transformers~\cite{Vaswani2017}, generative language modeling has seen huge progress. Large models, such as GPT-2~\cite{Radford2019}, shown generative language models to perform well across different tasks when fine-tuning them or in a few-shot setting~\cite{Brown2020, Smith2022}. Predictions of future performance promise that these models have the potential to even further improve their abilities~\cite{Kaplan2020, Hoffmann2022}.
While these models are evaluated on various tasks, we are not aware of any other systematic study of few-shot models on different code generation tasks.

\paragraph{Neural software analysis}

Our study is part of a larger stream of work on neural models of software~\cite{NeuralSoftwareAnalysis}.
An important question is how to embed code into a vector representation. Several approaches, e.g., based on AST paths~\cite{Alon2019}, control flow graphs~\cite{Wang2020a}, ASTs~\cite{Zhang2019}, and a combination of token sequences and a graph
representation of code~\cite{Hellendoorn2020} have been proposed.
The general-purpose generative model used here does not explicitly embed code into a vector representation, but instead relies on the ability of transformers~\cite{Vaswani2017} to reason about long-range dependencies.
Neural models of code address a wide range of problems, e.g.,
code completion~\cite{DBLP:journals/corr/abs-2004-05249,DBLP:conf/icse/KimZT021,Alon2019a},
type prediction~\cite{Hellendoorn2018,icse2019,fse2020,Wei2020},
program repair~\cite{Gupta2017,Dinella2020},
code search~\cite{Gu2018,Sachdev2018}, and
making predictions about code changes~\cite{Hoang2020,Brody2020}.
All the above approaches address a specific problem with a model designed for this problem.
Instead, our work studies how successful a general-purpose model is at competing with non-neural code manipulation tools.

\section{Conclusions}

This paper studies the strengths and limitations of few-shot, pre-trained language models for three popular code generation tasks.
By systematically comparing the recently proposed Codex model~\cite{Chen2021} against three traditionally built tools, we find that our model-based tools complement, are on par with, or even exceed the baseline tools.
At the same time, creating a new FSLM-based tool based on our methodology is relatively simple.
While our study shows promising results, we believe these are only first steps in applying few-shot learning to software engineering problems.

\bibliographystyle{ACM-Reference-Format}
\bibliography{references,referencesMP}

\end{document}